\documentclass[pra,twocolumn,showpacs,superscriptaddress,floatfix]{revtex4}
\usepackage[dvips]{graphicx,color}
\begin{document} 
  
\title{Formation of Multipartite Entanglement Using Random Quantum Gates}
  
\author{Yonatan Most, Yishai Shimoni and Ofer Biham} 
\affiliation{Racah Institute of Physics, The Hebrew University, 
Jerusalem 91904, Israel} 
  
\begin{abstract} 
    
The formation of multipartite quantum entanglement by repeated
operation of one and two qubit gates is examined. 
The resulting entanglement is evaluated using two measures:
the average bipartite entanglement and the Groverian measure. 
A comparison is made between two geometries of the quantum register: 
a one dimensional chain in which two-qubit gates apply only locally 
between nearest neighbors and a non-local geometry in which such gates 
may apply between any pair of qubits. More specifically, we use a 
combination of random single qubit rotations and a fixed two-qubit 
gate such as the controlled-phase gate.
It is found that in the non-local geometry the entanglement is 
generated at a higher rate. In both geometries, the Groverian 
measure converges to its asymptotic value more slowly than the 
average bipartite entanglement. These results are expected to have 
implications on different proposed geometries of future quantum
computers with local and non-local interactions between the qubits.
    
\end{abstract} 


\maketitle
  
\section{Introduction} 
\label{sec:intro}

Quantum algorithms such as
Shor's factoring algorithm 
\cite{Shor1994} 
and Grover's search
algorithm 
\cite{Grover1996,Grover1997a} 
exemplify  
the potential speedup offered by quantum computers. 
Although the origin of this speed-up is
not fully understood, 
there are indications that quantum entanglement
plays a crucial role 
\cite{Aharonov1996,Jozsa2003,Vidal2003}. 
Therefore, it is of interest to 
quantify the entanglement of the 
quantum states of multiple qubits 
that appear in quantum algorithms,
using suitable measures.
The special case of bipartite entanglement has been studied
extensively in recent years and 
was established as a resource for
quantum teleportation. 
The entanglement of pure bipartite
states is completely determined by their Schmidt coefficients. 
Thus, any measure of bipartite entanglement must be a function of these
coefficients. 
For example, the von Neumann entropy is 
expressed in terms of the eigenvalues of 
the reduced density matrix,
which are the squares
of the Schmidt coefficients.
For mixed bipartite states, 
several measures were proposed, 
namely the entanglement of
formation and the entanglement of distillation
\cite{Bennett1996a,Bennett1996b}. 
In particular, for states of two
qubits an exact formula for the entanglement of formation was obtained
\cite{Hill1997,Wootters1998}. 
The more general case of multipartite entanglement, 
in a register of $n>2$ qubits,
is not as well
understood, partly because no analogue of the Schmidt
decomposition was found for multipartite systems. 
Axiomatic considerations have provided a set of 
properties that entanglement
measures should satisfy
\cite{Vedral1997,Vedral1998,Vidal2000,Horodecki2000}. These
properties include the requirement that any entanglement measure
should vanish for product (or separable) states; it should be
invariant under local unitary operations and should not increase
as a result of any sequence of local operations complemented by
only classical communication between the parties. 
Quantities that satisfy these properties are called 
entanglement monotones. These
properties provide useful guidelines in the search
for entanglement measures for multipartite quantum states. 
Entanglement measures based on metric properties of the
Hilbert space 
\cite{Vedral1997,Vedral1997a,Vedral1998} 
and on polynomial invariants 
\cite{Barnum2001,Leifer2004} 
were proposed and shown to satisfy these requirements.
Although some measures have been studied extensively,
the connection between such measures and
the efficiency of quantum algorithms remains unclear. 

The common framework of quantum computation is based on a 
universal set of one and two qubit gates
\cite{Barenco1995b,Nielsen2000}.
Repeated operations of these gates enable to obtain any 
desired quantum state of the register.
To generate multipartite entanglement encompassing all
the qubits in a register, one applies a series of two 
qubit gates between pairs of qubits,
in addition to single qubit gates.
One can visualize this process as a network, in which
the qubits are represented by nodes and two qubit operations
are represented by edges that
connect the corresponding nodes.
Quantum entanglement is transitive in the sense that
if we apply a gate that entangles qubits $i$ and $j$
and then a gate that entangles qubits $j$ and $k$, 
this typically also gives rise to entanglement between 
qubits $i$ and $k$.
By entanglement between qubits $i$ and $k$ we mean
that any two parties, each including one of these qubits,
cannot be in a product state with each other. 
Thus, in order for the entanglement to encompass all the
qubits in the register, the nodes associated with
any pair of qubits must be connected, either directly or 
indirectly.

In recent years, several possible implementations 
of quantum computers have been proposed.
In some of these schemes the interactions between qubits
are non-local in the sense that two-qubit gates may apply
between any pair of qubits. 
Other schemes are based on a rigid geometry in which
only nearest neighbor qubits may interact.
In particular, they may be arranged in a one dimensional
chain, as in solid state quantum computers using quantum dots 
\cite{DiVincenzo2000}.
In a one-dimensional geometry, two qubit gates may apply 
on only $n-1$ or $n$ pairs of qubits (depending on the boundary conditions), 
out of the $n(n-1)/2$ possible pairs. 
It is interesting to examine to what extent the geometrical
restriction reduces the efficiency of the quantum computer.
This limitation was addressed in the context
of certain quantum algorithms 
\cite{Aharonov1997,Gottesman1999,Fowler2004a,Fowler2004b,Mottonen2005,Plenio2005}.

In this paper we consider the rate of formation of multipartite
entanglement in a register of $n$ qubits
by repeated operations of one and two-qubit gates.
Starting from a product state, we repeatedly apply a combination of random 
single qubit rotations and a fixed two-qubit gate on pairs of qubits.
Two geometries are considered: a non-local geometry in which any pair
of qubits may interact with each other and a 
one dimensional chain, in which two qubit
gates may apply only between nearest neighbor qubits.
The resulting entanglement is evaluated using two measures:
the average bipartite entanglement
\cite{Meyer2002,Brennen2003,Emerson2003}
and the Groverian measure
\cite{Biham2002}.
It is found that the non-local scheme is more effective,
namely it requires fewer steps to produce a certain level
of multipartite entanglement compared to the local scheme.

The paper is organized as follows.  
The entanglement measures used in this paper are 
briefly presented in Sec. II.
The entanglement generating schemes, 
with local and non-local geometries are 
shown in Sec. III. 
The simulations and results are presented in Sec. IV,
followed by a discussion in Sec. V and a 
summary in Sec. VI.

\section{Entanglement Measures of Multiple Qubits}
\label{sec:measures}

\subsection{The Average Bipartite Entanglement Measure}
  
Consider a quantum register of $n$ qubits
in a pure state 
$| \psi \rangle$. 
The bipartite entanglement between a given qubit 
and all the other qubits is given by
a single parameter 
\cite{Kraus2001}, 
namely, all the entanglement
monotones for such partition are the same up to a monotonic function. 
One of these monotones is the largest eigenvalue $P$ 
of the reduced density matrix, 
$\rho$, of the single qubit. 
This eigenvalue is also
the square of the largest Schmidt coefficient. 
Another monotone is 
${\rm Tr} (\rho^2) = P^2 + (1-P)^2$,
which is a monotonically increasing function of $P$
in the relevant interval 
$1/2 \le P \le 1$.
To evaluate the entanglement in a register of $n$ qubits we average
this measure over all $n$ choices of the single qubit,
where $\rho_i$, $i=1,2,\dots,n$ 
is the reduced density matrix of the $i$th
qubit, obtained by taking a partial trace over all the other qubits. 
After a suitable normalization and shift, 
one obtains the entanglement measure
\cite{Meyer2002,Brennen2003}
  
\begin{equation}
Q(\psi) = 2-\frac{2}{n} \sum_{i=1}^{n} {\rm Tr} \left (\rho_i^2 \right).
\label{eq:Q}
\end{equation}
  
\noindent
This measure was used in order 
to evaluate the
entanglement generated by 
repeated operations of
random gates
\cite{Emerson2003,Weinstein2005}. 
$Q(\psi)$ 
is an entanglement monotone only for the most refined 
partition, where each qubit is considered as a separate party.
It is thus commonly considered as
a measure of multipartite entanglement.

In fact, the measure 
$Q(\psi)$ 
essentially quantifies  
bipartite entanglement, which is averaged over
$n$ different partitions.
In each partition, one party consists of a single qubit
while the other party consists of $n-1$ qubits.
One may define other partitions of the
register into two parties, where one party includes $k$ qubits
and the other party includes $n-k$ qubits.
For a given value of $k$,
the entanglement measure is obtained by averaging 
Tr$(\rho^2)$ over all possible partitions of this type.
In a more general framework, the register can be partitioned
into any number of parties between 2 and $n$
\cite{Barnum2001,Shimoni2007}.

\subsection{The Groverian Measure}
\label{sec:groverian}
  
Grover's algorithm performs a search for a marked element $m$ in a
search space $D$ containing $N$ elements
\cite{Grover1996,Grover1997a}. 
We assume, for
convenience, that $N = 2^n$, where $n$ is an integer. This way, the
elements of $D$ can be represented by an $n$-qubit register $| x
\rangle = | x_1,x_2,\dots,x_n \rangle$, with the computational basis
states $| i \rangle$, $i=0,\dots,N-1$. The meaning of marking the
element $m$, is that there is a function $f: D \rightarrow \{0,1\}$,
such that $f=1$ for the marked element, and $f=0$ for the rest. To
solve this search problem on a classical computer one needs to
evaluate $f$ for each element, one by one, until the marked state is
found. 
In the worst case, this requires $N$ 
evaluations of $f$.
On a quantum computer, where $f$ can be
evaluated \emph{coherently}, 
Grover's algorithm, 
represented by the unitary operator $U_G$, 
can locate a marked element using only 
$O(\sqrt{N})$ coherent queries of $f$
\cite{Grover1996,Grover1997a}. The algorithm is based on 
quantum oracle, with the ability to recognize the
marked states
\cite{Grover1996,Grover1997a}.  
Starting with the equal superposition state, 
  
\begin{equation}
|\eta\rangle=\sum_{i=0}^{N-1}|i\rangle,
\end{equation}
  
\noindent 
and applying the operator $U_G$, one obtains the state
$U_G |\eta\rangle = |m\rangle + O({1}/{N})$,
which is then measured. 
The success probability of the algorithm is
almost unity. 
The adjoint equation takes the form
$\langle\eta| = \langle m|U_G + O({1}/{N})$.
If an arbitrary pure state, $|\psi\rangle$, is used as the initial
state instead of the state $| \eta \rangle$, the success probability
is reduced to
$P_s = |\langle m|U_G|\psi\rangle|^2 + O({1}/{N})$
or
$P_s=|\langle\eta|\psi\rangle|^2 + O({1}/{N})$.
The success probability is thus determined by the 
fidelity, between  
$| \psi \rangle$ 
and 
$| \eta \rangle$ 
\cite{Biham2002,Biham2003}.
  
Consider Grover's search algorithm, in which an arbitrary pure state
$| \psi \rangle$ is used as the initial state. Before applying the
operator $U_G$, there is a pre-processing stage in which arbitrary
local unitary operators $U_1$, $U_2$, $\dots$, $U_n$ are applied on
the $n$ qubits in the register. These operators are chosen such that
the success probability of the algorithm will be maximized. The
maximal success probability is thus given by
  
\begin{equation}
P_{\rm max}(\psi) = 
\max_{U_1,U_2,\dots,U_n}
|\langle m|U_G(U_1\otimes\dots\otimes U_n)|\psi\rangle|^2,
\label{eq:Pmax}
\end{equation}
  
\noindent
which can be re-written as
  
\begin{equation}
P_{\rm max}(\psi) = 
\max_{|\phi\rangle \in T}|\langle\phi|\psi\rangle|^2,
\label{eq:pmax}
\end{equation}
  
\noindent
where $T$ is the space of all tensor product states of the form
$|\phi\rangle = |\phi_1\rangle\otimes\dots\otimes|\phi_n\rangle$.
The Groverian measure is given by
$G(\psi) = \sqrt{1 - P_{\rm max}(\psi)}$
\cite{Biham2002}.
In the case of pure states, for which $G(\psi)$ is defined, it is
closely related to the entanglement measure introduced in
Refs.~\cite{Vedral1997,Vedral1997a,Vedral1998} for both pure and
mixed states and was shown to be an entanglement monotone. 
Based on these results, it was shown
\cite{Biham2002} that: 
$G(\psi) \geq 0$, with equality when $|\psi\rangle$
is a product state;
$G(\psi)$ is invariant under local unitary operations and
cannot be increased using local operations 
and classical communication.
Therefore, $G(\psi)$ is an entanglement monotone for pure states. A
related result was obtained in Ref.~\cite{Miyake2001}, where it was
shown that the evolution of the quantum state during the iteration
of Grover's algorithm corresponds to the shortest path in the Hilbert
space using a suitable metric.

The Groverian measure was used in order to evaluate the
entanglement generated by quantum algorithms such as Grover's
algorithm 
\cite{Shimoni2004} 
and Shor's algorithm
\cite{Shimoni2005}. 
It was also generalized to
the case of mixed states 
\cite{Shapira2006}
and to arbitrary partitions 
of the register
\cite{Shimoni2007}.
Here we use a different version of the Groverian measure,
referred to as the logarithmic Groverian measure, given
by

\begin{equation}  
G(\psi) = - \ln \left( P_{\rm max}(\psi) \right).
\label{eq:logGrov}
\end{equation}

\noindent
This is an entanglement monotone, because the logarithmic function
is monotonically increasing, and it vanishes for product states,
for which $P_{\rm max}=1$.
Unlike the measure  
introduced in Ref.
\cite{Biham2002},
which is restricted to the range of $[0,1)$,
the logarithmic Groverian measure may take values
in the range
$0 \le G(\psi) < \infty$.
This enables to better distinguish and compare between
highly entangled states which involve a 
large number of qubits. 

Furthermore, the logarithmic Groverian measure 
exhibits the additivity property described below.
Consider the state 
$| \psi \rangle = |\psi_{\rm A} \rangle  |\psi_{\rm B} \rangle$
where
$|\psi_{\rm A} \rangle$ 
and 
$|\psi_{\rm B} \rangle$
are pure states of
two different registers, of 
$n_{\rm A}$ 
and 
$n_{\rm B}$ 
qubits, respectively,
and
$n_{\rm A} + n_{\rm B} = n$. 
One can express the product state 
$| \phi \rangle$ of $n$ qubits
in the form
$| \phi \rangle = | \phi_{\rm A} \rangle | \phi_{\rm B} \rangle$,
where
$| \phi_{\rm A} \rangle$
and 
$| \phi_{\rm B} \rangle$
are product states of 
$n_{\rm A}$ 
and 
$n_{\rm B}$ 
qubits, respectively.
Since 
$| \psi \rangle$
is a tensor product of 
$|\psi_{\rm A} \rangle$
and   
$|\psi_{\rm B} \rangle$,
we obtain that
$\langle \phi | \psi \rangle = 
\langle \phi_{\rm A} | \psi_{\rm A} \rangle 
\langle \phi_{\rm B} | \psi_{\rm B} \rangle$. 
Thus, using Eq.
(\ref{eq:pmax})
we find that
$P_{\rm max}(\psi) = 
P_{\rm max}(\psi_{\rm A}) 
P_{\rm max}(\psi_{\rm B})$. 
As a result, 
the logarithmic Groverian measure 
satisfies

\begin{equation}
G(\psi) = G(\psi_{\rm A}) + G(\psi_{\rm B}).
\label{eq:additive}  
\end{equation}

\noindent
The additivity enables to
compare between the entanglement in registers which include
different numbers of qubits.

\section{The Entanglement Generating Scheme}
\label{sec:scheme}

To examine the rate of multipartite 
entanglement formation in quantum circuits
we consider a scheme for the production  
of pseudo-random states of $n$ qubits
\cite{Emerson2003}. 
The number of random amplitudes in a 
random state of $n$ qubits is exponential in $n$.
While the randomization of the state of each qubit alone 
requires only linear resources, 
the formation of entanglement between them is more costly.
Nevertheless, it
can be achieved by a combination of
random single qubit gates and a fixed
two-qubit gate, which apply repeatedly on pairs of qubits. 
It was shown that this
{\it rotate-entangle-rotate} 
scheme provides pseudo-random states with
only polynomial resources
\cite{Emerson2003,Oliveira2007}. 
The random single qubit rotation is obtained using

\begin{equation}
U_1 =
\left( {\begin{array}{*{20}c}
  {e^{2 \pi i\gamma _1 } \sqrt {1 - x} } & {e^{2 \pi i\gamma _2 } \sqrt x }  \\
  { - e^{ - 2 \pi i\gamma _2 } \sqrt x } & {e^{ -2 \pi i\gamma _1 } \sqrt {1 - x} }  \\
\end{array}} \right),
\end{equation}

\noindent
where at each step
$x$, $\gamma_1$ and $\gamma_2$
are drawn from a uniform distribution over the
unit interval $[0,1)$.
This parametrization provides operators
drawn uniformly from the Haar measure
\cite{Zyczkowski1994}. 
For the two-qubit gate, we use the
controlled-phase gate
\cite{Emerson2003} 

\begin{eqnarray}
U_2 = e^{ i \frac{\pi}{4} \sigma _z \otimes \sigma _z},
\end{eqnarray}

\noindent
expressed in the canonical
decomposition form
given in 
Ref.~\cite{Kraus2001}.
This is a sensible choice because this operator can
produce maximally entangled states. 

The initial state of the entanglement generating scheme
is a random product state, 
$|\psi_0 \rangle$,
of $n$-qubits,
where the state of each qubit is drawn from a uniform
distribution on the Bloch sphere.
Each iteration of this scheme consists of the following operations: 

\begin{enumerate}

\item 
Choose a random pair of qubits, $i$ and $j$,
from the $n$ qubits in the register. 
In the {\it non-local} scheme, any pair of qubits may be chosen
with equal probabilities. In the {\it local} scheme the qubits are
arranged in a one dimensional chain with periodic boundaries
and only nearest-neighbor pairs are chosen, 
with equal probabilites. 

\item 
Apply the controlled phase operator on 
qubits $i$ and $j$.

\item 
Apply random single qubit rotations,
drawn uniformly from the Haar measure
\cite{Zyczkowski1994}, 
on qubits $i$ and $j$.

\end{enumerate}

\noindent
The state obtained after $t$ iterations, or time steps,
is denoted by 
$| \psi_t \rangle$.
To examine the rate of entanglement formation
and to compare between the two schemes, 
we evaluate the following three functions of
$|\psi_t \rangle$: 
(a) the fidelity $F(\psi_0,\psi_t) = |\langle \psi_0 | \psi_t \rangle|$;
(b) the average bipartite entanglement $Q(\psi_t)$;
and (c) the logarithmic Groverian measure $G(\psi_t)$. 
Unlike $Q(\psi_t)$ and $G(\psi_t)$ which are entanglement measures,
$F(\psi_0,\psi_t)$ is not an entanglement monotone.
To put it on a common footing with the Groverian measure,
we define

\begin{equation}
K(\psi_0,\psi_t) = - \ln F(\psi_0,\psi_t),
\end{equation}

\noindent
which is a monotonically decreasing function of the
fidelity and takes values between zero and infinity. 
  
\section{Simulations and Results}

Each simulation of the entanglement-generating 
scheme creates a series of quantum states
$| \psi_t \rangle$, $t=1,2,\dots$.  
We examine the variation of
$K(\psi_0,\psi_t)$
$Q(\psi_t)$
and
$G(\psi_t)$
vs. $t$.
Due to the random nature of the operations, these
three functions strongly fluctuate when evaluated
for a single run of the entanglement generating scheme.
To reduce the noise and elucidate the systematic trends,
we perform a large number of runs using different initial
states $| \psi_0 \rangle$
and calculate the averages
$\langle K(\psi_0,\psi_t) \rangle$
$\langle Q(\psi_t) \rangle$
and
$\langle G(\psi_t) \rangle$
vs. $t$.

\begin{figure}
\includegraphics[width=\columnwidth,clip]{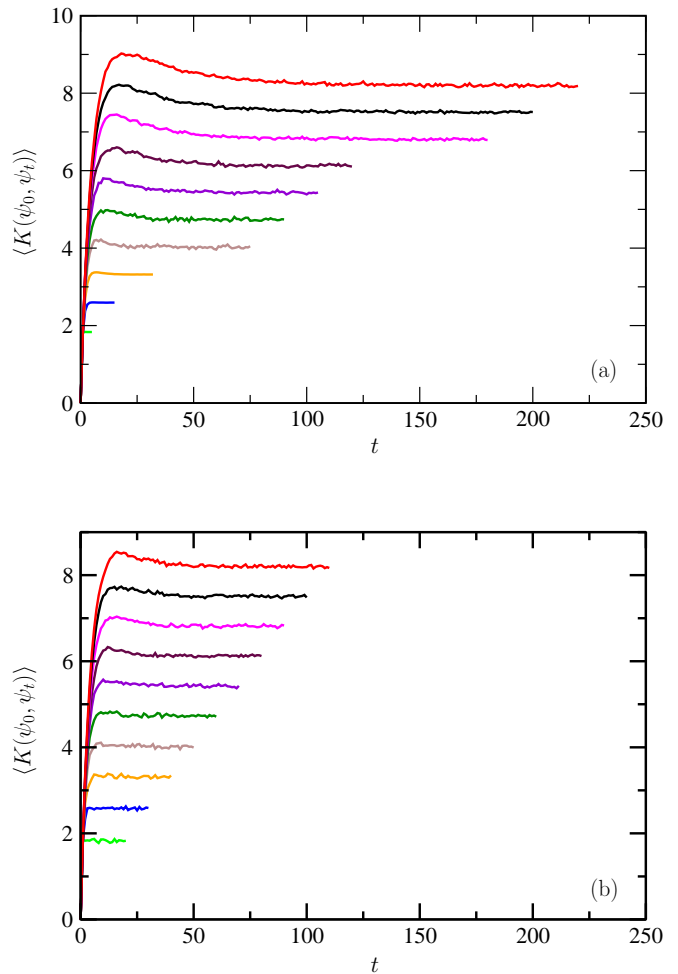}
\caption{
(Color online)
The average of 
$K(\psi_0,\psi_t) = - \ln F(\psi_0,\psi_t)$ 
vs. time 
in the local
scheme (a) and in the non-local scheme (b).
The number of qubits in the register is 
$n=2,3,\dots,11$ qubits (from bottom to top). 
}
\label{fig:1}
\end{figure}
  
In Fig.
\ref{fig:1}
we present 
$\langle K(\psi_0,\psi_t) \rangle$ vs. $t$,
averaged over 2000 realizations,
for the local scheme
[Fig. \ref{fig:1}(a)]
and for the non-local scheme
[Fig. \ref{fig:1}(b)]. 
The sharp increase of 
$\langle K(\psi_0,\psi_t) \rangle$ 
is explained by the fact that 
the fidelity is highly sensitive not only to two-qubit operations
but also to single-qubit operations.
Hence, the first few steps of the scheme
can decrease the fidelity substantially. 
To explain the overshoot,
notice that product states have smaller fidelity with each other
than entangled states, 
since the fidelity of two product
states is simply the multiplication of the fidelities of
the tensor components (all smaller then 1). 
In the beginning of the
scheme, the register is still nearly separable 
so its fidelity with the (also separable) initial state is
smaller than when it is completely entangled. 
After a sufficient number of steps, the state
of the register is uncorrelated with the input state, 
and 
$\langle K(\psi_0,\psi_t) \rangle$ 
saturates. 
As the number of qubits in the register increases, 
it takes more steps to bring
$\langle K(\psi_0,\psi_t) \rangle$ 
to saturation.
However, there is no significant 
difference between the two geometries.
  
\begin{figure}
\includegraphics[width=\columnwidth,clip]{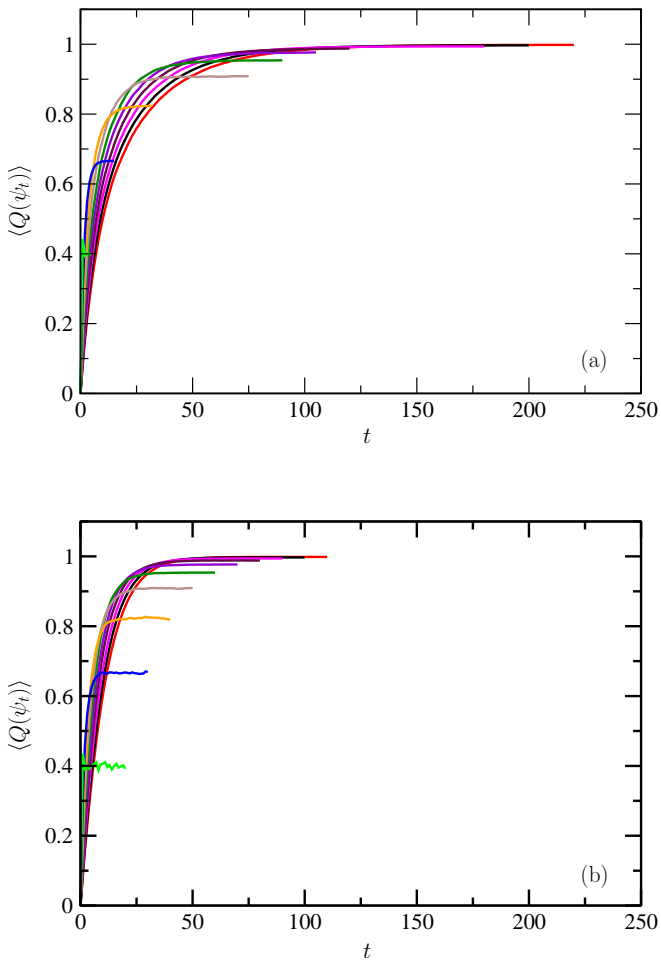}
\caption{
(Color online)
The average of the entanglement measure
$Q(\psi_t)$ vs. $t$
in the local
scheme (a) and in the non-local scheme (b), 
for a register with $2,3,\dots,11$ qubits,
from bottom to top. 
Clearly, the entanglement builds up faster in 
the non-local scheme.
}
\label{fig:2}
\end{figure}
  
\begin{figure}
\includegraphics[width=\columnwidth,clip]{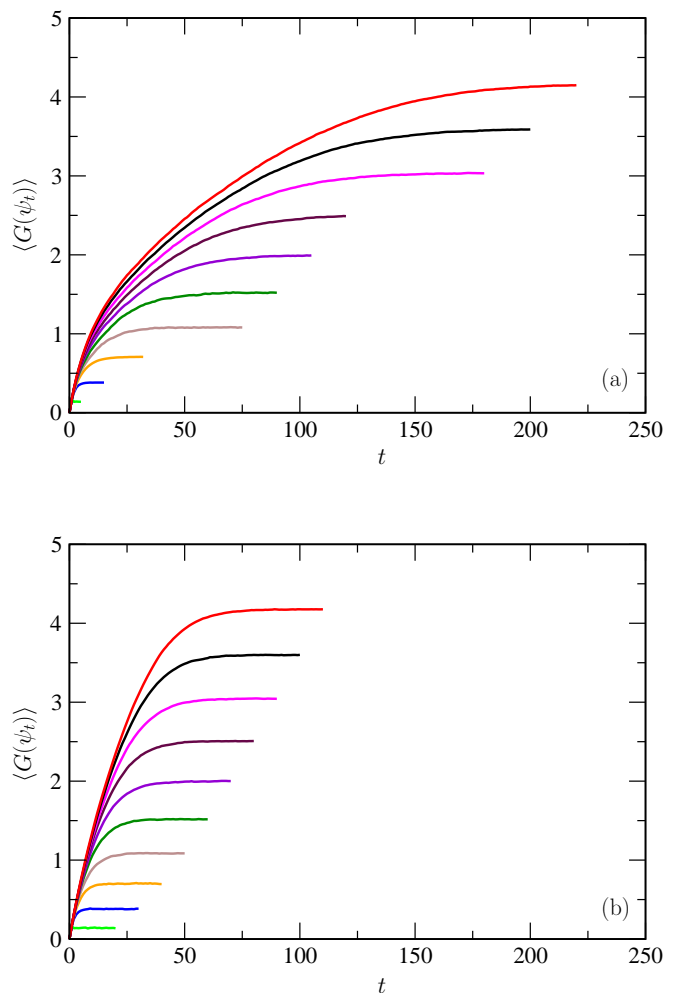}
\caption{
(Color online)
The average 
of the Groverian measure
$G(\psi_t)$ vs. $t$
in the local
scheme (a) and in the non-local scheme (b), 
for a register with $2,3,\dots,11$ qubits.
The Groverian measure increases faster in 
the non-local scheme.
}
\label{fig:3}
\end{figure}
  
The average bipartite entanglement, 
$\langle Q(\psi_t) \rangle$, 
is shown in 
Figs. \ref{fig:2}(a)
for the local scheme  
and in 
Fig. \ref{fig:2}(b)
for the non-local scheme. 
It increases more slowly than 
$\langle K(\psi_0,\psi_t) \rangle$. 
The asymptotic value of 
$\langle Q(\psi_t) \rangle$ 
is the same for both geometries, 
but it is reached much faster in the non-local geometry
than in the local, one dimensional geometry.
The results for the
one dimensional geometry are consistent with 
those presented in
Ref.~\cite{Emerson2003}.
The Groverian measure
$G(\psi_t)$ vs. $t$
is shown in 
Fig.
\ref{fig:3}(a) 
and
\ref{fig:3}(b),
for the local and non-local geometries, 
respectively.  
Clearly,
$\langle G(\psi_t) \rangle$ 
converges to its asymptotic value
more slowly
than $\langle Q(\psi) \rangle$.
The asymptotic value of 
$\langle G(\psi) \rangle$
is the same for both geometries
but it is reached much 
faster in the non-local geometry. 

\begin{figure}
\includegraphics[width=\columnwidth,clip]{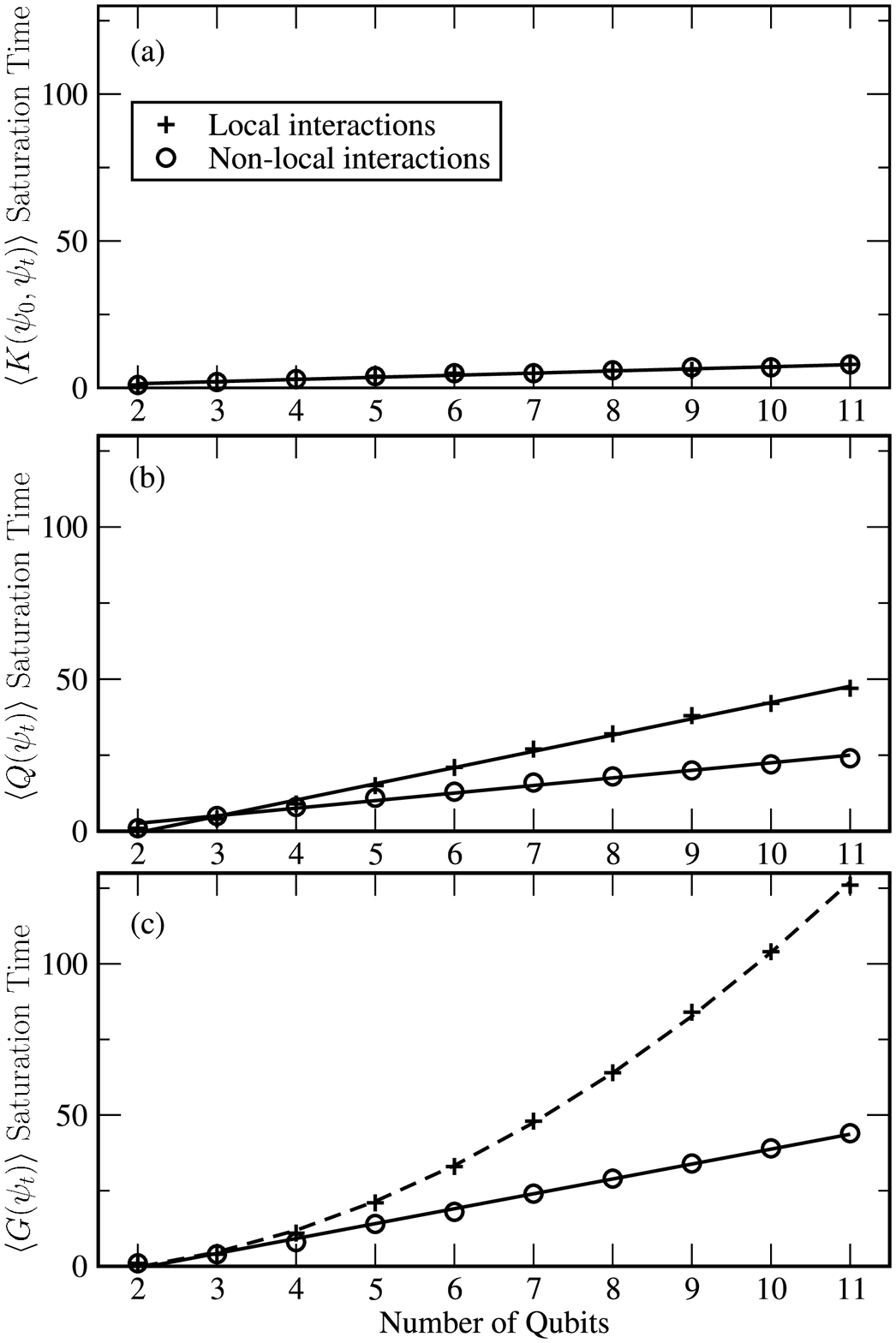}
\caption{
The number of steps required for the averages of 
$K(\psi_0,\psi_t)$ (a),
$Q(\psi_t)$ (b)
and
$G(\psi_t)$ (c)
to reach 90\% of their saturation values 
in the local 
($+$)
and in the non-local 
($\circ$)
schemes
vs. the number of qubits in the register.
While the results for
$K(\psi_0,\psi_t)$
are nearly identical in both schemes,
the time required for 
$Q(\psi_t)$ 
and
$G(\psi_t)$ 
to converge is shorter in the 
non-local scheme.
The solid lines are linear fits and the dashed
line is a quadratic fit.
}
\label{fig:4}
\end{figure}

The number of steps required for 
$\langle K(\psi_0,\psi_t) \rangle$,
$\langle Q(\psi_t) \rangle$,
and
$\langle G(\psi_t) \rangle$,
to reach
90\% of their saturation values, vs. $n$,
is shown in
Figs. \ref{fig:4}(a), 
\ref{fig:4}(b) 
and 
\ref{fig:4}(c),
respectively. 
The results for 
$\langle K(\psi_0,\psi_t) \rangle$
are almost identical in the local and
non-local schemes.
The time it takes
$\langle Q(\psi_t) \rangle$,
to reach
90\% of its saturation values appears to be linear in $n$,
for both geometries.
However, the slope is lower for the non-local geometry,
which means that the average bipartite entanglement builds
up more quickly when non-local interactions are allowed.
Similarly, 
in the non-local geometry
the saturation time of
$\langle G(\psi) \rangle$ 
is linear in $n$.
In the local geometry, the saturation time 
is longer and it
deviates from linear dependence on $n$
and is well fitted by a quadratic function of $n$.

In the analysis above we focused on the averages of
functions $K$, $Q$ and $G$.
Each data point was obtained by 
averaging over at least 2000 runs of the
entanglement forming procedure described above. 
These averages were taken from the distributions
$f(K)$, $f(Q)$ and $f(G)$ for the quantum states
$| \psi_t \rangle$
obtained after $t$ iterations of the 
procedure.
As $t$ increases, these distributions are found to approach
the distribution obtained for random states of the register
\cite{Emerson2003}. 
In Fig. 5 we present the probability densities $f(G)$
of the Groverian measure $G(\psi_t)$, for 
$t=10$ (dashed-dotted line), $20$ (dashed line) 
and $100$ (solid line) steps,
for a register with $n=8$ qubits.
As the number of steps increases, these distributions
approach the distribution $f(G)$ obtained for
pseudo-random states of the 8-qubit register
(dotted line).
  
\begin{figure}
\includegraphics[width=\columnwidth,clip]{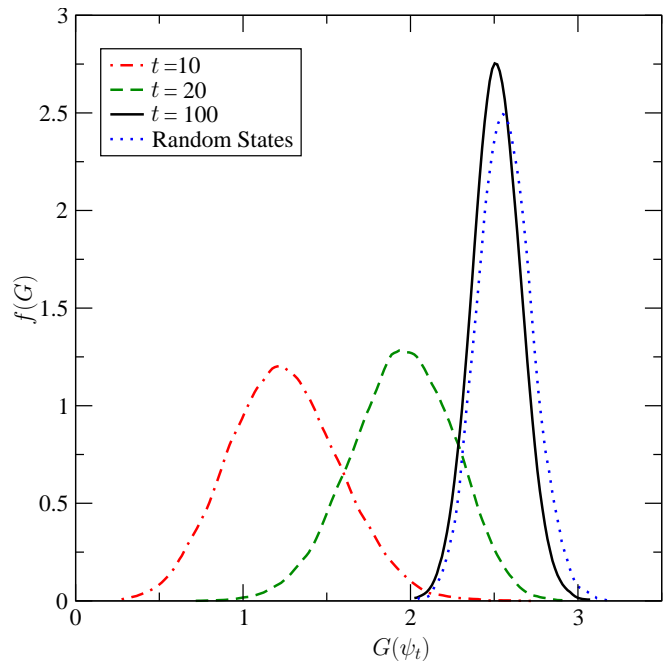}
\caption{
(Color online)
The probability density $f(G)$ of the Groverian
measure $G(\psi_t)$
for the states obtained after $t=10$ (dashed-dotted line),
20 (dahsed line) and 100 (solid line) 
iterations of the
entanglement forming procedure, for $n=8$ qubits.
As $t$ increases, the distribution converges to
the one obtained for pseudo-random states
(dotted line).
}
\label{fig:5}
\end{figure}
  
These results indicate that building up 
bipartite as well as multipartite entanglement,
using two-qubit gates is more efficient when
non-local interactions are allowed.
Furthermore, in both geometries it seems that 
multipartite entanglement, evaluated by
$G(\psi)$
builds up more slowly than the
bipartite-like entanglement,
quantified by $Q(\psi)$. 

\section{Discussion}
\label{sec:discussion}

Quantum algorithms are typically designed 
under the assumption that two-qubit operations are
possible between any pair of qubits.
However, some of the physical realizations of quantum computation
allow only local interactions between neighboring qubits
\cite{DiVincenzo2000}. 
In these realizations, non-local operations are achieved using
swap gates 
\cite{Aharonov1997},
which may involve a significant overhead
\cite{Schuch2003}.
To put the results of this paper in the context of 
known physical realizations,
we briefly review below
several such realizations and specify whether they are
based on local or non-local interactions.

Ion-trap quantum computing devices
\cite{Cirac1995,Kielpinski2003} 
consist of laser-cooled ions, 
which are electromagnetically confined
in ultra-high vacuum. 
The spin states of the ions are used as qubits.
Single-qubit gates are preformed  
by selectively applying electromagnetic fields on the
different ions.
Two-qubit gates are
realized through a global phonon state, which makes use of the
collective vibrational degree of freedom of the ions.
The ground state and the
first excited state of the phonon comprise a two-level system, which
is used as an additional qubit. 
Using carefully tuned laser-induced
transitions, two-qubit gates can be applied between each ion and the
global phonon. 
Specifically, a swap gate can be applied between them, 
enabling to apply two-qubit gates between any
pair of qubits. 
Ion traps thus provide a
non-local implementation of quantum computation. 
Linear optics
provides another non-local realization of quantum 
computation
\cite{Knill2001,Kok2007}.

Quantum computation can also be performed using solid state devices
\cite{Cerletti2005,Burkard2004}. 
Several proposals are comprised of
qubits based on superconducting Josephson junctions
\cite{Makhlin2001}. 
One such proposal uses the two states
of a superconducting single-charge box as a charge-qubit. 
The possibility of connecting all the qubits in parallel to a common
LC-oscillator mode was suggested as a way to apply two-qubit
gates 
\cite{Makhlin1999}. 
This parallel connection, together with
the ability to simultaneously control the Josephson coupling for
each qubit, allows two-qubit interaction between any pair of qubits,
giving rise to a non-local scheme. 
A different superconducting
scheme is based on Josephson flux qubits, 
which interact through magnetic induction. 
Unlike the charge-qubit scheme, 
the flux-qubit scheme is
a local one.
Another implementation is based on
electron spins as qubits, 
confined to quantum dots 
\cite{Loss1998}. 
Two-qubit gates are applied using a
time-dependent Heisenberg exchange coupling 
between adjacent dots, 
providing a local
scheme for quantum computation. 
Solid state realizations may be most suitable for scalable implementation
of quantum computations. Some of these realizations are based
on local interactions. 
Local interactions appear 
form multiple-qubit entanglement at a lower rate than non-local
interactions. However, the connection between the entanglement 
as quantified by the Groverian and related measures and the speedup 
offered by quantum algorithms is not yet clear. 
   
\section{Summary}
\label{sec:summary}

We have studied the 
formation of multipartite quantum entanglement by repeated
operation of one and two qubit gates.
The resulting entanglement was evaluated using 
the average bipartite and the Groverian measures. 
A comparison was made between two
geometries of the quantum register: 
a non-local geometry in which any pair of qubits may 
interact and a local geometry in which the qubits are
arranged in
a one dimensional chain,
where only nearest neighbor interactions are allowed. 
More specifically, we used a combination of 
random single qubit rotations
and a fixed two-qubit operation, 
namely the controlled phase gate.
We found that in this scheme, entanglement
is generated more quickly in the non-local geometry than in the
one-dimensional chain. 
Thus, non-local implementations of quantum computation are
expected to be more efficient in generating highly entangled
states.

\end{document}